\documentclass[fleqn,12pt,twoside]{article}
\usepackage{espcrc1}

\usepackage{graphicx}

\newcommand{\AmS}{{\protect\the\textfont2
  A\kern-.1667em\lower.5ex\hbox{M}\kern-.125emS}}
\newcommand{\bbox}[1]{\mbox{\boldmath{$#1$}}}

\def\G3H{$\Gamma^{\beta}_t$}
\def\AEXC{$\bbox{A}_{\rm EXC}$}
\def\AEXCNetal{$\bbox{A}_{\rm EXC}^{\rm NETAL}$}
\def\AEXCN{$\bbox{A}_{\rm EXC}^{\rm NSGK}$}
\def\signud{{$\sigma_{\nu d}$}}
\def\signudCC{{$\sigma_{\nu d}^{CC}$}}
\def\signudNC{{$\sigma_{\nu d}^{NC}$}}

\title{Neutrino-deuteron reactions at solar neutrino energies}

\author{S. Nakamura${}^a$, T. Sato${}^{a,b}$, 
  S. Ando${}^b$, T.-S. Park${}^b$, F. Myhrer${}^b$,
  V. Gudkov${}^b$\\ and K. Kubodera${}^b$\\ 
  ${}^a${Department of Physics, Osaka University, 
  Toyonaka, Osaka 560-0043, Japan}\\ 
${}^b${Department of Physics and Astronomy, 
University of South Carolina, Columbia, SC 29208, USA}}
       
\begin{document}

\maketitle

\begin{abstract}
In interpreting the SNO experiments, accurate estimates 
of the $\nu d$ reaction cross sections are of great importance.
In our recent work \cite{NETAL}, we have improved 
our previous calculation by updating some of its inputs
and by incorporating the results
of a recent effective-field-theoretical calculation.
The new cross sections are slightly ($\sim$1 \%) larger than 
the previously reported values. 
It is reasonable to assign 1\% uncertainty 
to the $\nu d$ cross sections 
reported here;
this error estimate does {\it not}
include radiative corrections.
\end{abstract}

\section{Introduction}

The establishment of the Sudbury Neutrino Observatory (SNO) 
has greatly increased the necessity of 
detailed theoretical studies of the neutrino-deuteron ($\nu d$)
reactions\cite{NETAL,NSGK,ando,kn,BCK}. 
At SNO, neutrino oscillations can be directly tested
by measuring the solar electron-neutrino flux using the
charged-current (CC) reaction ($\nu_e d\rightarrow e^-pp$)
and the total solar neutrino flux using 
the neutral-current (NC) reaction, 
$\nu_x d\rightarrow \nu_xpn$ ($x=e,\mu\
{\rm or}\ \tau$).
The recent SNO results\cite{ahmad} have given definitive
evidence for neutrino oscillations. 
To make the best use of the existing and future SNO data,
it is highly desirable to further improve
the theoretical estimates of the $\nu d$ cross sections.

A traditional approach for describing
nuclear electroweak processes consists in 
evaluating the contributions of 1-body impulse-approximation 
(IA) operators and 2-body exchange-current 
(EXC) operators defined in the Hilbert space of non-relativistic 
nuclear wave functions,
with the EXC terms derived from one-boson exchange diagrams.
We refer to this approach as SNPA 
(standard nuclear physics approach).
The calculations of
the $\nu d$ cross sections (\signud)\ based on SNPA
have been done by several authors \cite{kn};
the most recent work due to Nakamura {\it et al.}\ 
shall be referred to as NSGK \cite{NSGK}
and NETAL \cite{NETAL}.

An alternative approach,
effective field theory (EFT), has been applied 
to the $\nu d$ reaction by
Butler {\it et al.}(BCK)\cite{BCK}.
The EFT lagrangian used by BCK involves 
one unknown low-energy constant (LEC), $L_{1A}$,
which BCK adjusted to optimize fit to the \signud\ of NSGK.
After this fine-tuning, the results of BCK
were found to agree with those of NSGK within 1\% over the entire 
solar-neutrino energy region.
Very recently, Ando {\it et al.} have performed 
an EFT-motivated calculation of 
\signud\ \cite{ando} using a method called EFT*;
in this approach, originally proposed by Park et al.\cite{PKMR},
the electroweak transition operators 
are derived with a cut-off scheme EFT,
while the initial and final nuclear wave functions are 
obtained with the use of a
realistic phenomenological $NN$ potential.
The EFT* lagrangian contains an unknown LEC,
denoted by $\hat{d}^R$, which plays a role similar to $L_{1A}$ in BCK.
In EFT*, however, $\hat{d}^R$ can be determined
directly from the tritium $\beta$-decay rate \G3H,
which allows a parameter-free calculation
of the $\nu d$ cross sections.

We give here a concise account of the latest
SNPA calculation of the $\nu d$ reaction
carried out in NETAL \cite{NETAL}.
The main points of improvement over NSGK are as follows.
First, the updated value of the axial coupling constant $g_A$
is used.
Secondly, the treatment of 
the axial-vector exchange current (\AEXC) is improved. 
Since the $\nu d$ reaction in the solar-neutrino energy region
is dominated by the Gamow-Teller (GT) transition,
\AEXC\  is a crucial ingredient that controls the accuracy of
calculated cross sections.
Among the various terms in \AEXC,
the $\Delta$-excitation current is the most important one.
NETAL uses the $\Delta$-excitation current determined
in Ref. \cite{Schiavilla}
to reproduce the experimental value of \G3H.
Thirdly, NETAL employs the effective Fermi coupling constant 
$G_F^{\, \, \prime}$
that includes inner-radiative corrections instead of
$G_F$ used in NSGK. 
Furthermore, the stability of theoretical estimates
is investigated by comparing the new SNPA calculation 
with Ando {\it et al.}'s EFT* calculation \cite{ando}.

%
\section{Formalism}

The interaction hamiltonian ($H_W$) for a weak semileptonic process is
given by the product of the hadron current ($J_{\lambda}$) 
and the lepton current ($L^{\lambda}$) as
\begin{eqnarray}
H_W^{X} & = & \frac{G_F^X}{\sqrt{2}}\int d \mbox{\boldmath{$x$}}  [
J_{\lambda}^{X}( \mbox{\boldmath{$x$}})L^{\lambda}( \mbox{\boldmath{$x$}}) +
\mbox{h. c.}], 
\label{eq_Ham}
\end{eqnarray}
with  $X$=CC or NC, and
$G_F^{CC} (G_F^{NC})= G_F^{\, \, \prime}V_{ud} 
(G_F^{\, \prime})$, where
$G_F^{\, \prime}$ is the weak coupling constant determined from 
the $0^+$-$0^+$ $\beta$-decay rates,
and $V_{ud}$ is the K-M matrix element.
The hadron current is the sum of the vector ($V_\lambda$) 
and the axial  current ($A_\lambda$);
\begin{eqnarray}
J_{\lambda}^{CC}(\mbox{\boldmath{$x$}})
 & = & V_{\lambda}^{+}(\mbox{\boldmath{
$x$}}) +
A_{\lambda}^{+}(\mbox{\boldmath{$x$}})
\end{eqnarray}
for the CC reaction and
\begin{eqnarray}
J_{\lambda}^{NC}(\mbox{\boldmath{$x$}}) 
&=& (1-2 \sin^2 \theta_W )V_{\lambda}^{3
}(\mbox{\boldmath{$x$}}) +
A_{\lambda}^{3}(\mbox{\boldmath{$x$}}) 
-2 \sin^2 \theta_W V_{\lambda}^{s}(\mbox{
\boldmath{$x$}})
\label{eq_NC-current}
\end{eqnarray}
for the NC reaction. Here $\theta_W$ is the Weinberg angle.
The hadron current consists of 
the 1-body IA current and the 2-body EXC,
whose explicit forms can be found in Ref.\ [1].
Other details of the calculation 
including multipole expansion and
the cross section formula are given in Ref.\.[2].
For numerical results,
NETAL used the deuteron 
and the $NN$ scattering wave functions 
generated with the ANLV18 potential \cite{av18}.

\section{Results and Discussion}

\begin{figure}[h]
\begin{center}
\includegraphics[width=210pt]{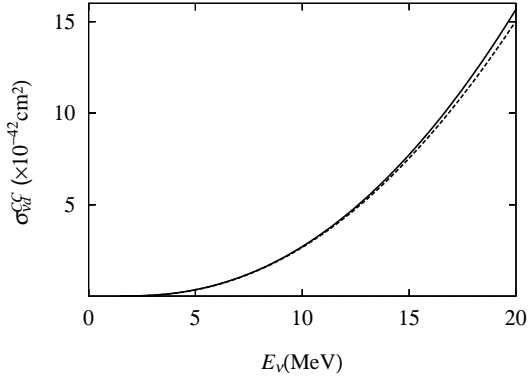}\\[-6mm]
\caption{Total cross sections for the CC reaction.
The solid- (dotted-) curve includes all (${}^1S_0$) partial waves of the
 final two-nucleon system.}
\label{fig_tot}
\end{center}
\end{figure}

The calculated total cross section for the CC reaction (\signudCC)\ is
shown in Fig.\ref{fig_tot}
as a function of the incident neutrino energy ($E_{\nu}$).
Although the cross section is dominated by the GT transition leading to 
the final ${}^1S_0$ $NN$ state,
the transitions leading to higher partial waves are not 
totally negligible; they account for 
1\% and 4\% of the reaction rate 
at $E_\nu \sim 10$ MeV and $E_\nu \sim 20$ MeV,
respectively.

\footnotesize
\begin{table}[h]
\begin{minipage}[t]{83mm}
  \begin{tabular}[t]{cccc}\hline
 $E_{\nu}$ (MeV) 
 & NETAL
 & NSGK
 & normalized \\\hline
  5& 1.019& 1.050& 1.020\\
 10& 1.023& 1.059& 1.024\\
 20& 1.029& 1.068& 1.030\\\hline
&&&\\[-2mm]
 \end{tabular}
 \caption{Contributions of EXC for the CC reaction. 
 The ratio,
  \signudCC(IA+EXC)/\signudCC(IA), is shown
for NETAL (2nd column)
and for NSGK (3rd column).
The fourth column corresponds to
the use of the ``normalized'' 
\AEXCN\  explained in the text .}
 \label{tab_exc}
\end{minipage}
\hspace{7mm}
\begin{minipage}[t]{70mm}
  \begin{tabular}[t]{ccc}\hline
 $E_{\nu}$ (MeV) 
 & $\nu_e d\rightarrow e^-pp$
 & $\nu d\rightarrow \nu pn$\\ \hline
 5& 1.003& 1.004\\
 10& 1.001& 1.003\\
 20& 0.998& 1.001\\\hline
&&\\[-2mm]
 \end{tabular}
 \caption{Comparison of SNPA and EFT*.
The ratio \signud(EFT*)/\signud(SNPA) 
is shown for the CC and NC reactions.}
 \label{tab_eft}
\end{minipage}
\end{table}
\normalsize
The ratio \signudCC(IA+EXC)/\signudCC(IA)
is shown in Table \ref{tab_exc} for both NETAL and NSGK.
In NETAL, the EXC contribution to \signudCC\
is $\sim$2\%,
which is $\sim$3\% smaller than the NSGK results.
The strength of \AEXCN\  was determined by analyzing the
$np \rightarrow \gamma d$ reaction and 
assuming a quark-model relation
between the axial vector and vector coupling constants
for the $N\Delta$ transition.
We note, however, that \G3H should give a much more direct
constraint on \AEXC\  than 
$\sigma(np\rightarrow d\gamma)$,
the latter being governed by the vector current.
One way to gauge the sensitivity of \signud\  to different choices of \AEXC\ 
is to normalize the strength of \AEXCN\ near threshold to
reproduce \AEXCNetal.
With this normalization applied,
the difference between the two EXC models is reduced to 0.2\%;
see the fourth column in Table 1.
Thus it is the overall strength of \AEXC\ 
that controls the low-energy $\nu d$ reactions;
\signud\ is insensitive to the detailed 
structure of \AEXC\  once EXC 
is adjusted to reproduce \G3H.

Comparison with an EFT* calculation \cite{ando}
provides a further test of the reliability of \signud\ obtained in NETAL.
It is sufficient to make this comparison
for the reaction rate leading to 
the final $^1S_0$ state.
Table \ref{tab_eft} shows the ratio  \signud(EFT*)/\signud(SNPA), from 
which we can conclude that
 SNPA and EFT* give identical results
within 1\% accuracy.
This agreement proves the robustness of the theoretical 
estimates of \signud\  obtained in NETAL.

With the improved inputs as described above,
NETAL have obtained \signud's
that are slightly larger ($\sim$1\%) than those of NSGK.
The \signud's in NETAL are considered to be reliable 
within 1\% accuracy.

Besides the absolute value of \signud, the ratio 
$R=$\signudNC$/$\signudCC\ is an important quantity 
for SNO experiments.
Comparing the $R$'s obtained in NETAL, NSGK and EFT*, 
we can assign $0.5$\% accuracy to the calculated $R$. 

The radiative corrections are expected to affect  \signud\ 
at the level of a few per cent.
NETAL only took into account a part of the radiative correction 
incorporated into $G_F^{\, \, \prime}$;
the most recent estimation of the remaining radiative corrections
can be found in Ref.\ \cite{kurylov}.

A new experimental value of $g_A$ has been presented
at this conference by Abele \cite{newgA}.
The reported value, $g_A$=1.274, is significantly
larger than the current PDG value used by NETAL.
We remark, however, that if the calculation of NETAL
is repeated with the use of this new value of $g_A$,
the resulting \signud's would be essentially unchanged.
Since the strength of \AEXC\  in NETAL
is determined to reproduce \G3H, an increase in $g_A$
is largely compensated by a decrease in \AEXC,
leaving \signud's essentially unaffected.

This work is supported in part by the Japan Society for the Promotion of 
Science, Grant No. (c) 12640273, and by the US National Science
Foundation, Grant No. PHY-9900756 and No. INT-9730847.

\end{document}